\documentclass[twocolumn,aps,superscriptaddress,showpacs]{revtex4}
\usepackage{amsmath,bm}
\usepackage{graphicx}

\begin{document}

\title{Double neutron/proton ratio of nucleon emissions in isotopic reaction
systems as a robust probe of nuclear symmetry energy}
\author{Bao-An Li}
\affiliation{Department of Chemistry and Physics, P.O. Box 419, Arkansas State
University, State University, Arkansas 72467-0419, USA}
\author{Lie-Wen Chen}
\affiliation{Institute of Theoretical Physics, Shanghai Jiao Tong University, Shanghai
200240, China}
\affiliation{Center of Theoretical Nuclear Physics, National Laboratory of Heavy Ion
Accelerator, Lanzhou 730000, China}
\author{Gao-Chan Yong}
\affiliation{Institute of Modern Physics, Chinese Academy of Science, Lanzhou 730000,
China}
\author{Wei Zuo}
\affiliation{Institute of Modern Physics, Chinese Academy of Science, Lanzhou 730000,
China}
\date{\today}

\begin{abstract}
The double neutron/proton ratio of nucleon emissions taken from two reaction
systems using four isotopes of the same element, namely, the neutron/proton
ratio in the neutron-rich system over that in the more symmetric system, has
the advantage of reducing systematically the influence of the Coulomb force
and the normally poor efficiencies of detecting low energy neutrons. The
double ratio thus suffers less systematic errors. Within the IBUU04
transport model the double neutron/proton ratio is shown to have about the
same sensitivity to the density dependence of nuclear symmetry energy as the
single neutron/proton ratio in the neutron-rich system involved. The double
neutron/proton ratio is therefore more useful for further constraining the
symmetry energy of neutron-rich matter. 
\end{abstract}

\maketitle





\section{Introduction}

The density dependence of nuclear symmetry energy $E_{\text{sym}}(\rho )$ is
still poorly known but very important for both nuclear physics and
astrophysics \cite{lat01,steiner,riatheory,ireview,ibook,baran05}. Heavy-ion
reactions induced by neutron-rich nuclei provide a unique opportunity to
constrain the symmetry energy in a broad density range. A number of
potentially useful probes of the symmetry energy have been proposed in the
literature mostly based on transport model simulations. Experimental data
from several dedicated experiments just started emerging. Some comparisons
between the available experimental data and transport model calculations have been
carried out recently. These studies have allowed us to place important
constraints on the density dependence of symmetry energy. For instance, by
using the free-space experimental nucleon-nucleon (NN) cross sections within the
transport model IBUU04 \cite{ibuu04}, a symmetry energy of $E_{\text{sym}%
}(\rho )\approx 31.6(\rho /\rho _{0})^{1.1}$ for densities less than $%
1.2\rho _{0}$ was extracted from the MSU data on isospin diffusion \cite%
{betty04,chen05}. While using in-medium NN cross sections
calculated within an effective-mass scaling approach \cite{li05}, a symmetry
energy of $E_{\text{sym}}(\rho )\approx 31.6(\rho /\rho _{0})^{0.69}$ was
found most acceptable in comparison with both the MSU isospin diffusion data
and the presently acceptable neutron-skin thickness in $^{208}$Pb \cite%
{steiner05,li05}. The currently existing isospin diffusion data alone can
not distinguish the above two forms of the symmetry energy within the
experimental error bars \cite{li05}. Thus complementary observables
sensitive to the $E_{\text{sym}}(\rho )$, more desirably, studies on
correlations of several such observables, are still very much needed to
further constrain the symmetry energy.

Because the symmetry potentials have opposite signs for neutrons and protons
and the fact that the symmetry potentials are generally smaller compared to
the isoscalar potential at the same density, most of the observables
proposed so far use differences or ratios of isospin multiplets of baryons,
mirror nuclei and mesons, such as, the neutron/proton ratio of nucleon
emissions \cite{li97}, neutron-proton differential flow \cite{dflow},
neutron-proton correlation function \cite{chen03}, $t$/$^{3}$He \cite%
{chen03a,zhang05}, $\pi ^{-}/\pi ^{+}$ \cite{li02,gai04,li05a,qli05a}, $\Sigma
^{-}/\Sigma ^{+}$ \cite{qli05b} and $K^{0}/K^{+}$ ratios \cite{ditoro05},
etc. Among these observables, the neutron/proton ratio of nucleon emissions
has probably the highest sensitivity to the symmetry energy. This is because
symmetry potentials act directly on nucleons and normally nucleon
emissions are rather abundant in typical heavy-ion reactions.
However, it is very challenging to measure some of these observables, especially
those involving neutrons. The measurement of neutrons, especially the low
energy ones, always suffers from low detection efficiencies even for the
most advanced neutron detectors. Therefore, observables involving neutrons
normally have large systematic errors. Moreover, for essentially all of these
observables, the Coulomb force on charged particles plays an important role.
It sometimes competes strongly with the symmetry potentials. One has to
disentangle carefully effects of the symmetry potentials from those due to the
Coulomb potentials. It is thus very desirable to find experimental observables which can
reduce the influence of both the Coulomb force and the systematic errors associated with neutrons.
The double neutron/proton ratio of nucleon emissions taken from two reaction
systems using four isotopes of the same element, namely, the neutron/proton ratio
in the neutron-rich system over that in the more symmetric system,
was recently proposed by Lynch et al.\cite{lynch} as a candidate of such an
observable. They have actually measured the double neutron/proton ratio
in central reactions of $^{124}$Sn$+^{124}$Sn and $^{112}$Sn$+^{112}$Sn
at a beam energy of $50$ MeV/nucleon at the National Superconducting Cyclotron
Laboratory\cite{lynch}. While the experimental data are currently
being finalized, we report here transport model analyses of the double
neutron/proton rations. Besides the above two reactions we also study the
double neutron/proton ratio in $^{132}$Sn$+^{124}$Sn and $^{112}$Sn$+^{112}$%
Sn reactions at $400$ MeV/nucleon. It is shown that the double
neutron/proton ratio has about the same sensitivity to the density
dependence of symmetry energy as the corresponding single ratio in the
respective neutron-rich system involved. Given the advantages of measuring
the double neutron/proton ratios over the single ones, the study of double neutron/proton
ratios will be more useful for further constraining the symmetry
energy of neutron-rich matter.

\section{A summary of the IBUU04 transport model}

\begin{figure}[tbh]
\includegraphics[scale=0.85]{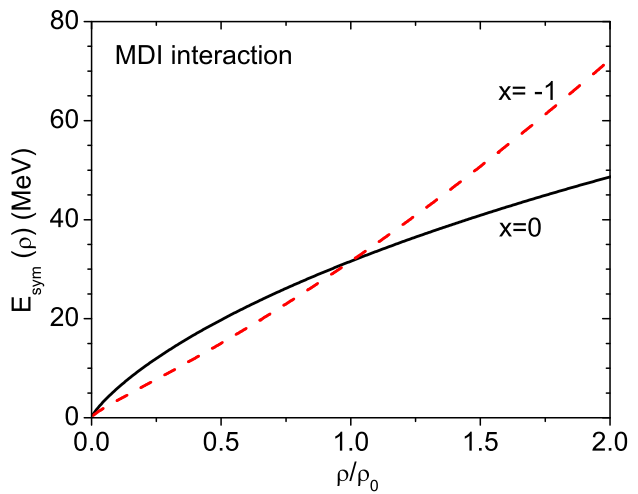}
\caption{{\protect\small (Color online) Symmetry energy as a function of
density for the MDI interaction with }$x=0${\protect\small \ and }$-1$%
{\protect\small .}}
\label{figure1}
\end{figure}

Our study is carried out using the IBUU04 version of an isospin and
momentum dependent transport model for nuclear reactions induced by
neutron-rich nuclei \cite{ibuu04}. For completeness and consistency we
outline here a few major features most relevant to the present study. More
details of the model can be found in Refs. \cite{ibuu04,li05}. The single
nucleon potential is one of the most important inputs to all transport
models for nuclear reactions. In the IBUU04 transport model, we use a single
nucleon potential derived within the Hartree-Fock approach using a modified
Gogny effective interaction (MDI) \cite{das03}, i.e.,
\begin{eqnarray}
U(\rho ,\delta ,\vec{p},\tau ,x) &=&A_{u}(x)\frac{\rho _{\tau ^{\prime }}}{%
\rho _{0}}+A_{l}(x)\frac{\rho _{\tau }}{\rho _{0}}  \notag \\
&+&B(\frac{\rho }{\rho _{0}})^{\sigma }(1-x\delta ^{2})-8\tau x\frac{B}{%
\sigma +1}\frac{\rho ^{\sigma -1}}{\rho _{0}^{\sigma }}\delta \rho _{\tau
^{\prime }}  \notag \\
&+&\frac{2C_{\tau ,\tau }}{\rho _{0}}\int d^{3}p^{\prime }\frac{f_{\tau }(%
\vec{r},\vec{p}^{\prime })}{1+(\vec{p}-\vec{p}^{\prime })^{2}/\Lambda ^{2}}
\notag \\
&+&\frac{2C_{\tau ,\tau ^{\prime }}}{\rho _{0}}\int d^{3}p^{\prime }\frac{%
f_{\tau ^{\prime }}(\vec{r},\vec{p}^{\prime })}{1+(\vec{p}-\vec{p}^{\prime
})^{2}/\Lambda ^{2}}.  \label{mdi}
\end{eqnarray}%
Here $\delta =(\rho _{n}-\rho _{p})/\rho $ is the isospin asymmetry of the
nuclear medium. In the above $\tau =1/2$ ($-1/2$) for neutrons (protons) and
$\tau \neq \tau ^{\prime }$; $\sigma =4/3$; $f_{\tau }(\vec{r},\vec{p})$ is
the phase space distribution function at coordinate $\vec{r}$ and momentum $%
\vec{p}$. The parameters $A_{u}(x),A_{l}(x),B,C_{\tau ,\tau },C_{\tau ,\tau
^{\prime }}$ and $\Lambda $ were obtained by fitting the momentum-dependence
of the $U(\rho ,\delta ,\vec{p},\tau ,x)$ to that predicted by the Gogny
Hartree-Fock and/or the Brueckner-Hartree-Fock (BHF) calculations \cite%
{bombaci}, the saturation properties of symmetric nuclear matter and the
symmetry energy of about $30$ MeV at normal nuclear matter density $\rho
_{0}=0.16$ fm$^{-3}$ \cite{das03}. The incompressibility $K_{0}$ of
symmetric nuclear matter at $\rho _{0}$ is set to be $211$ MeV consistent
with the latest conclusion from studying giant resonances \cite%
{k0data,pie04,colo04}. The parameters $A_{u}(x)$ and $A_{l}(x)$ depend on
the $x$ parameter according to
\begin{equation}
A_{u}(x)=-95.98-x\frac{2B}{\sigma +1},~~~~A_{l}(x)=-120.57+x\frac{2B}{\sigma
+1}.
\end{equation}%
The parameter $x$ can be adjusted to mimic predictions on the density
dependence of symmetry energy $E_{\text{sym}}(\rho )$ by microscopic and/or
phenomenological many-body theories. Shown in Fig.\ \ref{figure1} is the
density dependence of the symmetry energy for $x=0$ and $-1$. The recent
analyses of the MSU isospin diffusion data have allowed us to
constrain the $x$ parameter to be between these two values
for densities less than about $1.2\rho _{0}$\cite{li05}.
The corresponding symmetry energy can be parameterized as
$E_{\text{sym}}(\rho )\approx 31.6(\rho /\rho _{0})^{1.1}$
and $E_{\text{sym}}(\rho )\approx 31.6(\rho /\rho _{0})^{0.69}$ for $x=-1$
and $x=0$, respectively. The main purpose of this work is to investigate whether the
double neutron/proton ratio can help further narrow down the uncertainty of the
symmetry energy.

The last two terms in Eq. (\ref{mdi}) contain the momentum-dependence of the
single-particle potential. The momentum dependence of the symmetry potential
stems from the different interaction strength parameters $C_{\tau ,\tau
^{\prime }}$ and $C_{\tau ,\tau }$ for a nucleon of isospin $\tau $
interacting, respectively, with unlike and like nucleons in the background
fields. More specifically, we use $C_{unlike}=-103.4$ MeV and $%
C_{like}=-11.7 $ MeV. With these parameters, the isoscalar potential
estimated from $(U_{neutron}+U_{proton})/2$ agrees reasonably well with
predictions from the variational many-body theory \cite{wiringa}, the BHF
approach \cite{bombaci,zuo99,zuo} including three-body forces and the
Dirac-Brueckner-Hartree-Fock (DBHF) calculations \cite{sam05} in broad
ranges of density and momentum. For the MDI potential we used here, the
neutron-proton effective mass splitting due to the momentum dependence of
the symmetry potential is positive \cite{ibuu04}. This is consistent with
predictions of both the BHF and DBHF models \cite%
{bombaci,zuo99,zuo,sam05,fuchs05}.

The IBUU04 model can use either the free-space experimental NN cross
sections \cite{data} or the in-medium NN cross sections calculated using an
effective-mass scaling model consistent with the single particle potential
used \cite{li05}. In the present work the in-medium NN cross sections are
used.

\begin{figure}[tbh]
\includegraphics[scale=0.9]{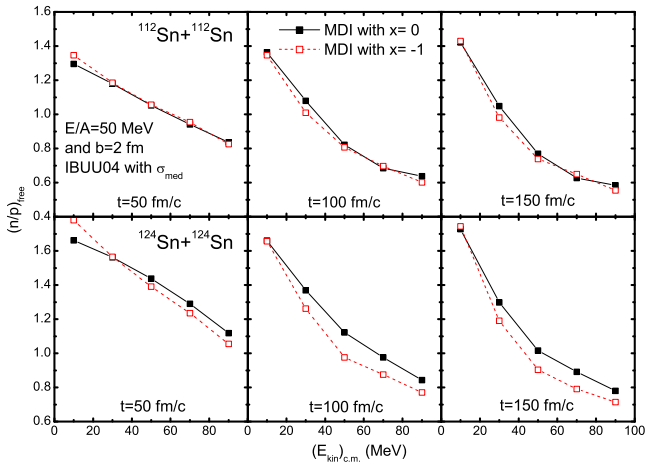}
\caption{{\protect\small (Color online) Time evolution of the neutron/proton
ratio of free nucleons as a function of kinetic energy for the reaction of }$%
^{124}${\protect\small Sn}$+^{124}${\protect\small Sn (lower windows) and }$%
^{112}${\protect\small Sn}$+^{112}${\protect\small Sn\ (upper windows) at }$%
50${\protect\small \ MeV/nucleon and an impact parameter of }$2$%
{\protect\small \ fm, respectively. The calculations are done with the MDI
interaction of }$x=0${\protect\small \ (filled square) and }$x=-1$%
{\protect\small \ (open square).}}
\label{figure2}
\end{figure}

\section{Results and Discussions}
We present and discuss here the double neutron/proton ratios of free nucleons in
central and mid-central reactions using three Sn isotopes at a beam energy of
$50$ MeV/nucleon and $400$ MeV/nucleon, respectively.
Free nucleons are identified as those having local baryon densities less than $\rho _{0}/8$.
For a comparison and as an example, we first study the single neutron/proton ratio of free
nucleons for $^{124}$Sn$+^{124}$Sn and $^{112}$Sn$+^{112}$Sn reactions at a beam energy 
of $50$ MeV/nucleon and an impact
parameter of $2$ fm, respectively. Shown in Fig.\ \ref{figure2} are the time
evolutions of the single neutron/proton ratios versus the nucleon kinetic energy in the
c.m.s frame of the respective reaction. It is seen that the neutron/proton ratio
becomes stable after about $100$ fm/c. As one expects, the neutron/proton ratio
in the neutron-richer system is more sensitive to the symmetry energy,
especially for fast nucleons. With the softer symmetry energy of $x=0$, the
symmetry energy and the magnitude of the symmetry potential are higher at
sub-saturation densities compared to the case with $x=-1$. At
supra-saturation densities, however, it is just the opposite as shown in
Fig.\ \ref{figure1}. In the above two reactions the maximum density reached
is about $1.2\rho _{0}$\cite{chen05}. One thus expects to see a higher
neutron/proton ratio of free nucleons with the softer symmetry energy of $x=0$ due to the
stronger repulsive (attractive) symmetry potential for neutrons (protons). For
the more symmetric system $^{112}$Sn$+^{112}$Sn, effects of the symmetry
energy are negligible because of the small isospin asymmetry in the system.
The rise of the neutron/proton ratio at low energies in both systems
is due to the Coulomb force which pushes protons away from the
center of mass of the reaction. These features are all consistent with those
found in an earlier study using a momentum-independent transport model \cite{li97}.
Unlike the results in Ref. \cite{li97}, however, the observed symmetry
energy effect is only about $10\%$ to $15\%$ even for the most energetic nucleons
in the $^{124}$Sn$+^{124}$Sn reaction. This is understandable since
the symmetry energy used here has already been severely constrained by the recent
isospin diffusion data. In ref.\cite{li97}, however, a much wider uncertainty
range between approximately $30(\rho/\rho_0)^{0.5}$ and $30(\rho/\rho_0)^2$
was used for the symmetry energy.
\begin{figure}[tbh]
\includegraphics[scale=1.2]{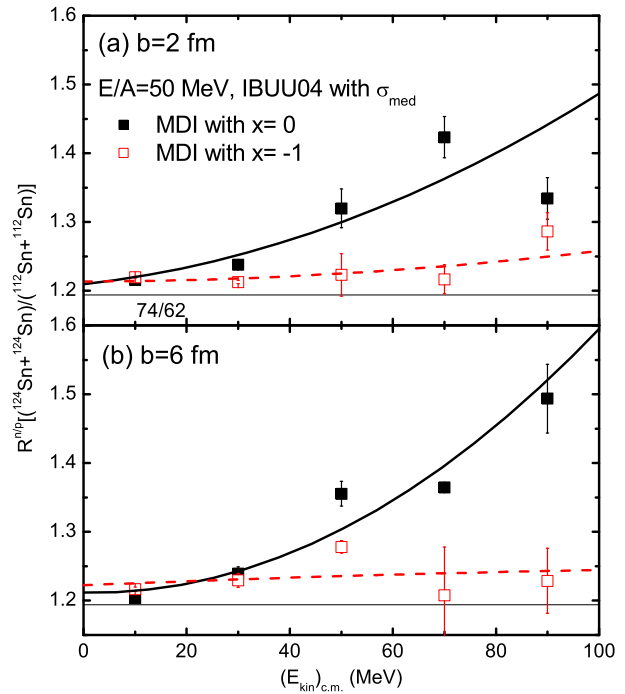}
\caption{{\protect\small (Color online) The double neutron/proton ratio of
free nucleons taken from the reactions of }$^{124}${\protect\small Sn}$%
+^{124}${\protect\small Sn and }$^{112}${\protect\small Sn}$+^{112}$%
{\protect\small Sn at }$50${\protect\small \ MeV/nucleon and an impact
parameter of }$2${\protect\small \ fm (upper window) and }$6${\protect\small %
\ fm (lower window), respectively.}}
\label{figure3}
\end{figure}

We now turn to the double neutron/proton ratio of free nucleons.
Shown in Fig.\ \ref{figure3} are the double neutron/proton ratios
calculated for the $^{124}$Sn$+^{124}$Sn and $^{112}$Sn$+^{112}$Sn
reactions at a beam energy of $50$ MeV/nucleon and an impact
parameter of $2$ fm (upper window) and $6$ fm (lower window),
respectively. As a reference, a straight line at $74/62$
corresponding to the double neutron/proton ratio of the entrance
channel is also drawn. Below the pion production threshold,
statistically, one would expect the double neutron/proton ratio of
nucleon emissions to be a constant close to this value neglecting
effects due to both the Coulomb and symmetry potentials. Indeed,
the observed double neutron/proton ratios especially at low
kinetic energies with $x=-1$ at both impact parameters are almost
constants just slightly above the straight line at $74/62$. The
fact that the double neutron/proton ratio is slightly higher than
$74/62$ even with $x=-1$ (which corresponds to a weaker symmetry
potential at sub-saturation densities compared to the case with
x=0) is due to the appreciable repulsive/attractive symmetry
potential on neutrons/protons in the $^{124}$Sn$+^{124}$Sn
reaction. For the double ratios in the two reactions involving
isotopes of the same element one expects that the Coulomb effects
are largely cancelled out. More energetic nucleons have gone
through denser regions of the reactions, effects of the symmetry
potentials on them are thus higher especially in the case with $x=0$.
Therefore, the double neutron/proton ratios increase when the $x$
parameter is changed from $x=-1$ to $x=0$, especially for
energetic nucleons.

At both impact parameters, effects of the symmetry energy are about $10\%-15\%$
changing from $x=-1$ to $x=0$ for the energetic nucleons which are mostly from
pre-equilibrium emissions. The observed
sensitivity to the symmetry energy is about the same as the single
neutron/proton ratio shown in Fig.\ 2. It should also be mentioned that
since the neutron/proton ratio at kinetic energies less than about $50$ MeV
is rather insensitive to the symmetry energy in the reactions at a beam
energy of $50$ MeV/nucleon, neutron detectors with a threshold energy of $50$
MeV is sufficient for the study discussed here. However, as we shall discuss
in the following, with reactions at beam energies above the pion production
threshold even the low energy neutrons are useful.

\begin{figure}[tbh]
\includegraphics[scale=0.85]{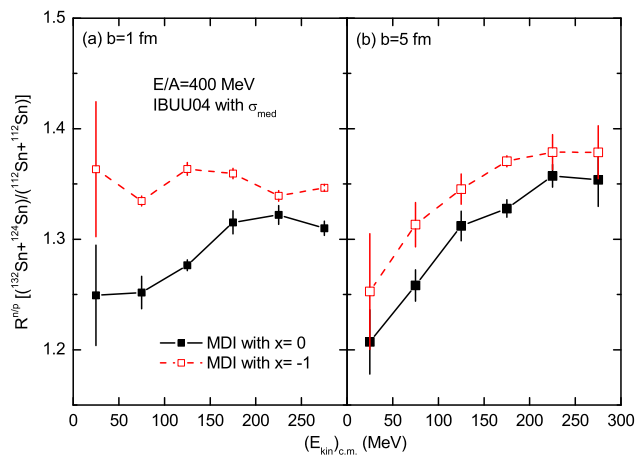}
\caption{{\protect\small (Color online) The double neutron/proton ratio of
free nucleons taken from the reactions of }$^{132}${\protect\small Sn}$%
+^{124}${\protect\small Sn and }$^{112}${\protect\small Sn}$+^{112}$%
{\protect\small Sn at }$400${\protect\small \ MeV/nucleon and an impact
parameter of }$1${\protect\small \ fm (left window) and }$5${\protect\small %
\ fm (right window), respectively.}}
\label{figure4}
\end{figure}

Moving to beam energies above the pion production threshold, the reference
line at $74/62$ is no longer useful. In fact, the $\pi ^{-}/\pi ^{+}$ ratio
itself is a promising probe of the symmetry energy at high
densities\cite{li02,gai04,li05a,qli05a}. Although the symmetry energy
at supra-saturation densities is currently not constrained by any experimental data, for this
study, we keep using the two parameters $x=0$ and $x=-1$ to be consistent
with the symmetry energy used for sub-saturation densities. Shown in Fig.\ %
\ref{figure4} are the double neutron/proton ratios from the reactions
of $^{132}$Sn$+^{124}$Sn and $^{112}$Sn$+^{112}$Sn at a beam energy of
$400$ MeV/nucleon and an impact parameter of $1$ fm (left window)
and $5$ fm (right window), respectively. At both impact parameters,
effects of the symmetry energy are about $5\%-10\%$ changing from the case with
$x=0$ to $x=-1$. One notices here that the low energy nucleons are having
the largest sensitivity to the variation of the symmetry energy for such
high energy heavy-ion collisions. In fact, the neutron/proton
ratio of midrapidity nucleons which have gone through the high density phase
of the reaction are known to be most sensitive to the symmetry energy \cite{li05b}.
Compared to the results at the beam energy of $50$ MeV/nucleon, it is interesting to see a
clear turnover in the dependence of the double neutron/proton ratio on the $x
$ parameter, namely the double ratio is lower at $50$ MeV/nucleon but higher
at $400$ MeV/nucleon with $x=-1$ than that with $x=0$. The maximum density
reached at the beam energy of $50$ and $400$ MeV/nucleon is
about $1.2\rho _{0}$ and $2\rho _{0}$ \cite{li05b}, respectively.
The turnover clearly indicates that the double neutron/proton ratio
reflects closely the density dependence of the symmetry energy as shown in Fig. \ref{figure1}.
This observation also indicates that systematic studies of the double neutron/proton ratio over a
broad beam energy range will be important for mapping out the density
dependence of the symmetry energy. It is useful to mention that the
double $\pi ^{-}/\pi ^{+}$ ratio in the reactions at $400$ MeV/nucleon has also been
examined. There is some indication that the double $\pi ^{-}/\pi ^{+}$ ratio
for intermediate energy pions is more sensitive to the symmetry energy than
the single $\pi ^{-}/\pi ^{+}$ ratio in the $^{132}$Sn$+^{124}$Sn reaction.
However, unlike the double neutron/proton ratio in the same ensemble of
events, the double $\pi ^{-}/\pi ^{+}$ ratio has much larger statistical
errors mostly because of the poor statistics of the single $\pi ^{-}/\pi ^{+}$ ratio in
the $^{112}$Sn$+^{112}$Sn reaction. Studies on the double $\pi ^{-}/\pi ^{+}$
ratio with significant more events are in progress.

\section{Summary}

In summary, within the transport model IBUU04 we investigated the double
neutron/proton ratio of nucleon emissions taken from two reaction systems
using three Sn isotopes at the beam energy of $50$ MeV/nucleon and
$400$ MeV/nucleon, respectively. It is found that the double neutron/proton
ratio has about the same sensitivity to
the density dependence of symmetry energy as the single neutron/proton ratio
in the more neutron-rich system of the two reactions. Since the double
neutron/proton ratio has the advantage of reducing systematically the
influence of the Coulomb force and has smaller systematic errors, it is more
useful than the single neutron/proton ratio of nucleon emissions for further
constraining the symmetry energy of neutron-rich matter.

We would like to thank Zoran Basrak, W.G. Lynch, Wolfgang Trautmann and M.B.
Tsang for helpful discussions. The work of B.A. Li is supported in part by
the US National Science Foundation under Grant No. PHY-0354572, PHY0456890
and the NASA-Arkansas Space Grants Consortium Award ASU15154. The work of
L.W. Chen is supported in part by the National Natural Science Foundation of
China under Grant No. 10105008 and 10575071. The work of G.C. Yong and W.
Zuo is supported in part by the Chinese Academy of Science Knowledge
Innovation Project (KJCX2-SW-N02), Major State Basic Research Development
Program (G2000077400), the National Natural Science Foundation of China
(10575119, 10235030) and the Chinese Ministry of Science and Technology (2002CAB00200).%
%


\end{document}